\documentclass[11pt,twoside]{article}

\usepackage{asp2006}
\usepackage{epsf}
\usepackage{psfig}

\begin{document}

\title{The Hobby-Eberly Telescope Dark Energy Experiment (HETDEX): Description and Early Pilot Survey Results}

\author{
Gary J.\ Hill, \altaffilmark{1} 
Karl Gebhardt, \altaffilmark{1}
Eiichiro Komatsu, \altaffilmark{1}
Niv Drory, \altaffilmark{2}
Phillip J.\ MacQueen, \altaffilmark{1}
Josh Adams, \altaffilmark{1}
Guillermo A.\ Blanc, \altaffilmark{1}
Ralf Koehler, \altaffilmark{2}
Marc Rafal, \altaffilmark{1}
Martin M. Roth, \altaffilmark{3}
Andreas Kelz, \altaffilmark{3}
Caryl Gronwall, \altaffilmark{4}
Robin Ciardullo, \altaffilmark{4}
and Donald P.\ Schneider \altaffilmark{4}
}

\altaffiltext{1}{McDonald Observatory \& Department of Astronomy,
    University of Texas at Austin, 1 University Station, Austin, TX 78712, USA}
\altaffiltext{2}{Max-Planck-Institut f\"ur Extraterrestrische Physik,
  Giessenbachstrasse, D-85748 Garching b. M\"unchen, Germany}
\altaffiltext{3}{Astrophysikalisches Institut Potsdam,
  An der Sternwarte 16, 14482 Potsdam, Germany}
\altaffiltext{4}{Department of Astronomy, Pennsylvania State University, 525 Davey Lab,
  University Park, PA 16802, USA}

\begin{abstract} 
The Hobby-Eberly Telescope Dark Energy Experiment (HETDEX) will outfit the 10 m HET with a new wide field and an 
array of 150 integral-field spectrographs to survey a 420 deg$^2$ area in the 
north Galactic cap. 
Each fiber-coupled unit spectrograph will cover 350-550 nm, simultaneously.
This instrument, called VIRUS, 
will produce $\sim$34,000 spectra per exposure, 
and will open up the emission-line universe to large surveys for the first time.
The survey will detect 0.8 million 
Lyman-alpha emitting (LAE) galaxies with 1.9$<$$z$$<$3.5 and more than 
a million [OII] emitting galaxies with $z$$<$0.5. 

The 3-D map of LAE galaxies in 9 cubic Gpc volume will be used to measure
the expansion history at this early 
epoch using baryonic acoustic oscillations and the shape of the power spectrum. 
The aim of HETDEX is to provide a direct detection of dark energy at z$\sim$3.
The measurement will constrain the evolution of dark energy and will also
provide 0.1\%-level accuracy on the curvature of the Universe,
ten times better than current.

The prototype of the VIRUS unit spectrograph (VIRUS-P) is a powerful 
instrument in its own right.  Used on the McDonald 2.7~m, 
it covers the largest area of any integral field spectrograph, 
and reaches wavelengths down to 340 nm. 
VIRUS-P is being used for a pilot survey to better measure the properties of 
LAE galaxies in support of HETDEX.  We report initial results from this 
survey.

\end{abstract}


\section{The HETDEX Approach to the Problem of Dark Energy}

Progress in understanding the physical nature of dark energy
will require precision measurements of the expansion history of 
the Universe over the redshift range 0$<$$z$$<$4.  
In order  to make progress towards this goal, 
very significant surveys involving new facilities are required. 
Two approaches are being pursued: to refine the 
accuracy of the measurement at low redshift, including constraints on 
possible evolution of the dark energy
equation of state, or to constrain dark energy evolution, directly, 
through observations at high redshift.

HETDEX
\Citep{h2004b} has the goal of providing percent-level constraints on the
expansion history of the universe (the Hubble parameter
$H(z)$ and angular diameter distance $D_A(z)$) over redshifts
$z$=1.9 to 3.5. 
HETDEX will use a combination of baryonic acoustic oscillations
\Citep[e.g.][]{se2007,ksg2007}) 
and power spectrum shape
information to provide at least a 3-$\sigma$
direct detection of dark energy over these redshifts, even in the event
that dark energy is a cosmological constant. To achieve this, 
HETDEX needs an accuracy of 0.9\% on $H(z)$ at
$z$=2.8. This level of accuracy requires a volume of 9
Gpc$^3$ with a density of tracers $\sim$$10^{-4}$ objects per
Mpc$^3$, which can be achieved by surveying 420
deg$^2$ over 1.9$<$$z$$<$3.5 with 
0.8 million Lyman-$\alpha$ emitting (LAE) galaxies. 
LAEs have high number density and are easily detected with integral field 
spectroscopy (as shown in Sec. 3).
In addition to a direct detection of dark energy, a 0.9\% measurement of
$D_A$ at $z$=$3$ from HETDEX will determine
curvature to 0.1--0.2\% \Citep{k2006} 
a factor of ten better than currently known. 
Nearly all of the other
dark energy missions require some knowledge of curvature to disentangle
the dark energy contribution. This is because at low redshifts, dark
energy and curvature provide the same expansion signature.  

\section{The HET Wide Field Upgrade and VIRUS}

HETDEX has three aspects: an upgrade to the HET, the instrument VIRUS,
and the observing and analysis campaign. The HET has pioneered an
optical design 
with a fixed spherical primary and a tracker to follow the 
motions of objects. 
The design is particularly effective 
for surveys. The wide field upgrade is needed to
increase the science field-of-view from 4\arcmin\ to 22\arcmin. This
entails a new corrector, tracker, and
instrument package \Citep{HETWFU}. 

The survey necessary to realize the desired constraints for HETDEX is
a significant undertaking, and requires a telescope/spectrograph
combination which can acquire the data an order of magnitude faster
than current spectrographs. 
For such surveys to be tractable, a new
approach to instrumentation is needed. 
Industrial replication \Citep{HillMacQueen02}, 
promises the multiplex gains required for the
next generation of instruments on very large telescopes.  

The {\bf V}isible {\bf I}ntegral-field {\bf R}eplicable {\bf U}nit 
{\bf S}pectrograph (VIRUS) 
consists of 150 integral-field spectrographs. 
Each VIRUS module has a fiber-coupled IFU feeding a pair of simple 
unit spectrographs.    
The VIRUS design is described in more detail in 
\Citet{VIRUS04,VIRUS06a,VIRUS06b,VIRUS08a,VIRUS08b}.
Each VIRUS unit is fed by 224 fibers that each
cover 1.8 arcsec$^2$ on the sky. 
The fibers feeding a two-unit module are arrayed in a 
50$\times$50 square arcseconds
IFU with a 1$/$3 fill-factor. 
A dither pattern of three exposures fills in the area.
The spectral resolution is
5.7\AA, with coverage of 350--550~nm. 
The optical design is simple, using three reflective
and two refractive elements. 
With dielectric reflective coatings optimized for the 
wavelength range, high throughput is obtained.  
The full VIRUS array will simultaneously obtain 33,600 spectra 
with 12~million resolution elements. 
The IFUs are arrayed within the 
22\arcmin~field of view of the upgraded HET with $\sim$1$/$7 fill factor, 
sufficent to detect the required density of LAEs for HETDEX. 
Development is proceeding with the prototype (VIRUS-P, \Citet{VIRUS08b}), 
deployed in October 2006, 
and pre-production prototype where value engineering is being used to 
reduce the cost for production
\Citep{VIRUS08a}. 

\begin{figure}[!ht]
\plotone{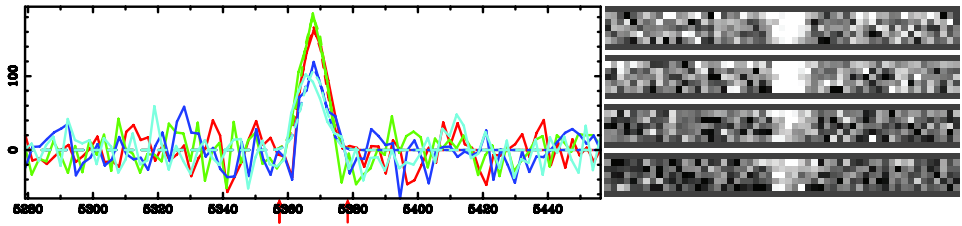}
\caption{Example spectrum of a z=3.415 LAE in the COSMOS field from
  HETDEX pilot survey data obtained on the McDonald 2.7~m with VIRUS-P.
   The LAE is seen in four separate dithered fiber positions
   (shown on the right),
   and combined for a strong detection.}
\end{figure}

\section{Properties of LAE Galaxies from the HETDEX Pilot Survey}

The HETDEX observing
campaign will take about 1400 hours on the HET (dependent on the still
poorly understood properties of LAEs), which we estimate to
require about 3 years to complete. 
Accurate measurements of the LAE luminosity function at $z=3$ and 4 are
now available from narrow-band imaging surveys and some
spectroscopy to confirm candidates 
\Citep[e.g.][and references elsewhere in these proceedings]{g2007}.
These studies currently probe insufficient volume to measure the
correlation function and bias of the LAEs, and they do not extend to the
lower redshifts probed by HETDEX.
To understand the properties of LAEs in more
detail, to measure the contaminating fraction of low redshift
galaxies, and to demonstrate the power of integral field spectroscopy
for selecting large samples, we have started a pilot survey using
VIRUS-P on the McDonald 2.7~m.  We are targeting the
MUNICS-deep, COSMOS, and GOODS-N fields, where deep imaging and other
supporting data exist, allowing us to characterize the properties of
detected emission-line objects.
VIRUS-P is ideal for this investigation, as the IFU covers 3.5
arcmin$^2$ when fed at f/3.65 f-ratio.  The individual fibers are
4.1\arcsec\ diameter, but the sensitivity is adequate to detect
significant numbers of LAEs, even on a 2~m class telescope. 
In comparison to narrow-band imaging surveys,
VIRUS-P covers a wide redshift range, down to
the uncharacterized $z$$=$$2$ epoch, and it surveys volume quickly.

To date in 80 nights, 
we have covered 115 arcmin$^2$ or 7.5x10$^5$ cubic Mpc co-moving volume,
reaching 5-6x10$^{-17}$ erg/cm$^2$/s line flux (5-$\sigma$) in 2 hours exposure.
Figure 1 presents an example LAE spectrum.
Analysis of 40 arcmin$^2$ in COSMOS yields 99 emission-line objects
(45 secure LAEs, 43 low redshift galaxies, one AGN, and 10 sources that are yet to be 
confirmed as LAEs).
Separation of LAEs from [OII] emitters is based on the deep imaging 
available for the targeted fields.
An imaging survey of moderate depth (AB$\sim$25) is needed to separate the two
populations, based on emission-line equivalent width \Citep{g2007}.

We are detecting significant numbers of LAEs at $z$$\sim$2.
Comparison of the number detected with the expected redshift distribution,
based on the measured throughput of the instrument and a non-evolving 
$z$$\sim$3 luminosity function, already gives
hints about LAE evolution. 
An increase in the galaxy luminosity function derived from UV and IR 
continuum measurements
of about a factor of two is seen between redshifts $z$$\sim$4 and 2
\Citep{r2008}.
Current data from the pilot survey 
are consistent with this trend, but analysis of the full
dataset will be required for significant results.
The pilot survey has proven the approach of wide-area integral field 
spectroscopy for HETDEX and has tested the data reduction pipeline.
In another year observing we expect to refine the properties of LAEs
needed for the full survey.


With 2$/$3 of the budget raised, HETDEX officially started in September
2007.  
Following the 
Science Requirments Review in June 2007 and Preliminary Design Review
in April 2008, the project is now in the 3-year build-phase.

\acknowledgements 

HETDEX is a collaboration between The University of
Texas at Austin, Max-Planck-Institut f\"ur Extraterrestrische Physik, 
Universit\"at-Sternwarte Munich, Astrophysikalisches
Institut Potsdam, Pennsylvania State University, Texas A\&M
University, Stanford University,  
and Georg-August Universitaet G\"ottingen.
We thank the members of the Science Requirements Review panel
(R. Bacon, G. Bernstein (Chair), E. Kolb, S. Rawlings)
and the PDR panel
(B. Bigelow, G. Chanon, R. Kurz, A. Russell (Chair), R. Sharples)
for their valuable input that has helped to guide the project.
We thank the Directors 
D. Lambert, R. Bender, \& M. Steinmetz,
and staff of McDonald observatory, MPE, and AIP, 
and the staff of the HET, 
for their support of HETDEX.
In particular
J. Murphy, F. Grupp, J. Booth, M.P. Smith, J. Good, R. Savage, B. Vattiat,
M. Shetrone, S. Odewahn, G. Damm, \& P. Palunas.
The VIRUS prototype was funded by George and Cynthia Mitchell.
The HETDEX pilot survey is supported by the Texas Advanced
Research Program under Grant No. 003658-0005-2006



\begin{thebibliography}{}


\bibitem[Booth et al.\ (2006)]{HETWFU}
Booth, J.~A., MacQueen, P.~J., Good, J.~G., Wesley, G.~L., Segura, P.~R.,
Palunas, P., Hill, G.~J. \& Calder, R.~E., 2006, Proc. SPIE, 6267, 128

\bibitem[Gronwall et al.\ (2007)]{g2007}
Gronwall, C. et al.\ 2007, ApJ, 667, 79

\bibitem[Hill \& MacQueen (2002)]{HillMacQueen02}
Hill, G.~J., \& MacQueen, P.~J., 2002, Proc. SPIE, 4836, 306

\bibitem[Hill et al.\ (2004a)]{VIRUS04}
Hill, G.~J., MacQueen, P.~J., Tejada, C., Cobos, P.~J.,  Palunas, P,
Gebhardt, K., \& Drory, N., 2004a, Proc. SPIE, 5492, 25

\bibitem[Hill et al.\ (2004b)]{h2004b}
Hill, G.~J., Gebhardt, K., Komatsu, E., \& MacQueen, P.~J., 2004b,
in AIP Conf. Proc., 743,
The Mitchell Symposium on Observational Cosmology, 
ed. R. E. Allen, D. V. Nanopoulos, \& C. N. Pope
(New York: AIP), 224

\bibitem[Hill et al.\ (2006a)]{VIRUS06a}
Hill, G.~J., MacQueen, P.~J., Tufts, J.~R., Kelz, A., Roth, M.~M.,  
Altmann, W., Segura, P., Smith, M., Gebhardt, K., \& Palunas, P., 2006a, 
Proc. SPIE, 6269, 93

\bibitem[Hill et al.\ (2006b)]{VIRUS06b}
Hill, G.~J., MacQueen, P.~J., Palunas, P., Kelz, A., Roth, M.~M.,
Gebhardt, K., \& Grupp, F., 2006b, New Astronomy Reviews, 50, 378

\bibitem[Hill et al.\ (2008a)]{VIRUS08a}
Hill, G.~J., MacQueen, P.~J., Palunas, P.,  \& Shetrone, M.
2008a, Proc. SPIE, 7014-5

\bibitem[Hill et al.\ (2008b)]{VIRUS08b}
Hill, G.~J., et al.,~2008b, Proc. SPIE, 7014-257 

\bibitem[knox (2006)]{k2006}
Knox, L., 2006, PhysRevD, 73, 023503

\bibitem[Koehler, Schuecker, \& Gebhardt (2007)]{ksg2007}
Koehler, R., Schuecker, P., \& Gebhardt, K., 2007, A\&A, 462, 7

\bibitem[Reddy et al.\ (2008)]{r2008}
Reddy, N.~A., Steidel, C.~C., Pettini, M., Adelberger, K.~L., Shapley, A.~E.,
Erb, D.~K., \& Dickinson, M., 2008, ApJS, 175, 48

\bibitem[Seo \& Eisenstein (2007)]{se2007}
Seo, H.-J., \& Eisenstein, D. 2007, ApJ, 665, 14  


\end{thebibliography}
\end{document}